\documentclass[aps,prl,twocolumn,superscriptaddress]{revtex4}

\usepackage{epsfig}
\usepackage{amsmath}
\usepackage{graphicx}
\usepackage[mathcal]{eucal}

\def\eV{\hbox{eV}}
\def\GeV{\hbox{GeV}}

\def\vev#1{\mathopen\langle #1\mathclose\rangle }
\def\chibar{{\overline{\chi} }}
\def\phibar{{\overline{\phi} }}

\begin{document}

\title{Fermion Mass Hierarchy and Proton Stability from
Non-anomalous $U(1)_F$ in SUSY $SU(5)$}

\author{Mu-Chun Chen}\email{muchunc@uci.edu}
\affiliation{Department of Physics \& Astronomy,
 University of California, Irvine, CA 92697, USA}

\author{D.~R. Timothy Jones}\email{drtj@liverpool.ac.uk}
\affiliation{Department of Mathematical Sciences, University of
Liverpool, Liverpool L69 3BX, UK}
 
\author{Arvind Rajaraman}\email{arajaram@uci.edu}
\affiliation{Department of Physics \& Astronomy,
 University of California, Irvine, CA 92697, USA}
 
\author{Hai-Bo Yu}\email{haiboy@uci.edu}
\affiliation{Department of Physics \& Astronomy,
 University of California, Irvine, CA 92697, USA}

\date{April, 2008}

\preprint{\vbox{\hbox{UCI-TR-2007-54}}}
\preprint{\vbox{\hbox{LTH 790}}}

\begin{abstract}
We present a realistic supersymmetric $SU(5)$ model combined with 
a non-anomalous $U(1)_{F}$ symmetry. 
We find a set of $U(1)_F$ charges which 
automatically lead to the realistic mass hierarchy and mixing
patterns for quarks, leptons and neutrinos.
All gauge anomalies, including the $[U(1)_{F}]^{3}$ anomaly, are
cancelled in our model without invoking the Green-Schwarz
mechanism or having exotic fields. 
Proton decay mediated by dimension 5 operators is automatically
suppressed in our model, because the scale set by the largest 
right-handed neutrino mass is much less than the GUT scale. 
\end{abstract}

\maketitle
\section{Introduction}
The fermion mass hierarchy and mixings are some of the least 
understood aspects of the Standard Model (SM). Many attempts have
been made to understand the large disparity among the masses and
mixing angles~\cite{Chen:2003zv}.
One 
approach is the Froggatt-Nielsen (FN) mechanism \cite{Froggatt:1978nt},
where an $U(1)_F$ family symmetry is introduced under which the SM
fermions are charged.  The $U(1)_{F}$ symmetry is broken by the
vacuum expectation value  of a SM-singlet scalar field $\phi$
whose $U(1)_{F}$ charge, without loss of generality, is normalized
to be -1. Fermion masses are generated by the operators
\begin{eqnarray}\label{fn}
Y_{ij}\left(\frac{\phi}{\Lambda}\right)^{|q_i+q_j+q_H|}
\overline{\Psi}_i\Psi_jH
\; ,
\end{eqnarray}
if $( q_{i} + q_{j} + q_{H} ) $ is a positive integer,  
and $\phi$ 
is replaced by $\phi^{\dagger}$ if $(q_{i} + q_{j} +
q_{H})$ is a negative integer. Here $i,j$ are generation indices,
and $q_i, \, q_j$ and $q_H$ are respectively the $U(1)_F$ charges
of the fermions $\overline{\Psi}_i$, $\Psi_j$ and the SM Higgs doublet $H$.
The parameter $\Lambda$ is the cutoff scale of the $U(1)_{F}$
symmetry. Upon breaking the $U(1)_{F}$ symmetry,
 the effective Yukawa couplings can be written as
\begin{eqnarray}
Y^{eff}_{ij}=Y_{ij}\lambda^{|q_i+q_j+q_H|} \; ,
\end{eqnarray}
where $\lambda = \left< \phi \right> / \Lambda $ if $( q_{i} +
q_{j} + q_{H} ) $ is a positive integer, and $\lambda = \left<
\phi^{\dagger}\right> / \Lambda $ if $(q_{i} + q_{j} + q_{H})$ is a
negative integer. By having appropriate $U(1)_{F}$ charges for
various fermions, the realistic masses and mixing patterns can be
accommodated, with $\lambda$ being smaller than unity and
$Y_{ij}\sim {\cal O}(1)$. If the model is supersymmetric, which is 
the case in our model, the couplings to the $\phi^{\dagger}$ are not
allowed, since the superpotential must be  holomorphic.
However, in order to ensure
D-flatness, 
a $\overline{\phi}$ field, which carries charge +1, must be
introduced, with a vacuum expectation value close to that of $\phi$. 
In this case, $\lambda = \left<
\overline{\phi}\right> / \Lambda $ if $(q_{i} + q_{j} + q_{H})$ is a
negative integer. Note that as a result the contributions of $\phi,\phibar$ 
to gauge anomalies cancel. 

Models based on a global $U(1)_{F}$ symmetry have been constructed
before (see references in \cite{Chen:2003zv}). However, as any
global symmetry is broken by quantum gravity effects, one inevitably
has to promote the $U(1)_{F}$ symmetry to be a gauge symmetry. In
this case, the $U(1)_{F}$ charges of the fields are constrained by
the anomaly cancellation conditions.
Most attempts~\cite{Irges:1998ax,Dreiner:2007vp,Duque:2008ah} so far have focussed on
an {\it anomalous} $U(1)_F$ symmetry, in which the
mixed anomalies are cancelled by the Green-Schwarz
mechanism~\cite{Green:1984sg}, while the $[U(1)_{F}]^{3}$ anomaly 
is not addressed. The $[U(1)_{F}]^{3}$ anomaly can be 
cancelled by introducing exotic matter fields charged under
$U(1)_F$~\cite{Bijnens:1987ff}. However, these models often involve a
rather large number of exotic particles whose role is merely to
cancel the anomalies~\cite{Kang:2004ix}.

Here we pursue an alternative scenario~\cite{Chen:2006hn} where the theory is
anomaly-free, without invoking the Green-Schwarz mechanism or
exotic particles other than the right-handed neutrinos,
which are required 
for neutrino masses. (We do have the FN pair of SM singlets $\phi,\phibar$ 
described above,  
and we will find it necessary to introduce another such oppositely-charged 
pair, but these do not contribute to gauge anomalies.)   
We propose a SUSY $SU(5)$ model, combined with
a {\it non-anomalous} $U(1)_{F}$,  in the presence of right-handed
neutrinos.  We find a set
of $U(1)_{F}$ charges that satisfy {\it all} the anomaly cancellation conditions,
including the $[U(1)_{F}]^{3}$ anomaly.
Note that these charges have to be rational numbers in order for the model to be
embedded into a simple group in the UV completed theory;
it is thus highly non-trivial for these
solutions to exist. We show that these $U(1)_{F}$ charges give rise to
realistic  quark
and charged lepton masses and mixing patterns. 

The charges we find also lead to a FN mixing pattern for the Dirac 
neutrino Yukawa matrix, $Y_{\nu}$; however to  accommodate the
right-handed neutrino masses  required for the see-saw mechanism we need
to introduce a second pair of  SM singlet fields $(\chi,\chibar)$ with
charges $\mp \frac{5}{9}$. This leads to a realistic  texture for the
light neutrino mass matrix, and if we assume that the observed atmospheric 
neutrino $(\hbox{mass})^2$ difference sets the scale of the heaviest 
neutrino mass, it follows that $\vev{\chi},\vev{\chibar}\sim 10^{11}\GeV$. 

Furthermore we then find that the dimension 4 R-parity violating 
operators ${\bf10}_i {\bf\overline5}_j{\bf\overline5}_k$,  
are forbidden, 
and moreover the dimension 5 operators which mediate proton decay  are
automatically suppressed without additional assumptions or fine-tuning
of parameters, because they arise  from higher dimensional operators in
the effective field theory involving powers  of $\chi,\chibar$. This
solves a major problem with the usual minimal SUSY $SU(5)$ GUT theories,
and allows our model to be viable.

\section{The Model}
In $SU(5)$, the three generations of matter fields are unified
into ${\bf\overline 5}_i$ and ${\bf 10}_i$ representations, where
$i = 1, 2, 3$ is the generation index. Under $U(1)_F$, 
${\bf\overline 5}_i$ and ${\bf 10}_i$ have charges
$q_{f_{i}}$ and $q_{t_{i}}$, respectively. We also introduce
right-handed neutrinos, the number of which 
is a free parameter.  If the type-I
seesaw~\cite{seesaw} is the mechanism that gives rise to light
neutrino masses, we need at least two right-handed neutrinos to
accommodate the current neutrino oscillation data; 
we will take three right-handed neutrinos,
$N_i$, which are $SU(5)$ singlets and carry $U(1)_F$ charges,
$q_{n_{i}}$. 

To generate realistic fermion mass hierarchy utilising the
FN mechanism
and to cancel the gauge anomalies, it turns out that two
conjugate pairs of $\bf 5$ and $\bf\overline 5$ Higgses are
required (this will be clear once the $U(1)_F$ charges are presented),
which we denote as ${\bf
5}_{H_{1}}$, ${\bf\overline{5}}_{H_{1}}$, ${\bf 5}_{H_{2}}$ and
${\bf\overline{5}}_{H_{2}}$.  The $U(1)_{F}$ charges of 
${\bf5}_{H_{1}}$ and ${\bf5}_{H_{2}}$ are $q_{H_{1}}$ and
$q_{H_{2}}$ respectively. In addition, we need a $\bf
24$-dim Higgs to break $SU(5)$ to the SM gauge group. We
take this Higgs to be neutral under the $U(1)_{F}$ symmetry.
The $U(1)_{F}$ symmetry is broken spontaneously by the vacuum
expectation values of the $SU(5)$ singlets, $\phi$ and
$\overline\phi$, whose $U(1)_F$ charges are normalized to $-1$ and
$+1$, respectively. Note that $\langle\phi\rangle -
\langle\overline{\phi}\rangle << \Lambda$, as required by  D-flatness.

There are three anomaly cancellation conditions that have to be
satisfied: the $[SU(5)]^{2} U(1)_{F}$, gravitation-$U(1)_{F}$ and
$[U(1)_{F}]^{3}$ anomalies. Since the $\bf 24$-dim Higgs is
neutral under the $U(1)_{F}$ symmetry, it does not contribute to
these anomalies. As the $\bf 5$-dim Higgses and $\phi$ all appear
in conjugate pairs, they do not contribute either. Therefore, only
the matter fields, ${\bf\overline5}_i,~{\bf10}_i$ and right-handed
neutrinos $N_i$, contribute to the anomalies. To cancel the
anomalies, their $U(1)_{F}$ charges must satisfy
\begin{eqnarray}
\frac{1}{2} \sum_{i} q_{f_{i}} + \frac{3}{2} \sum_{i} q_{t_{i}}=0\label{ano1} \; ,\\
5 \sum_{i}q_{fi} + 10 \sum_{i}q_{t_{i}} + \sum_{i}q_{n_{i}}=0\label{ano2} \; ,\\
5 \sum_{i}q^3_{fi} + 10 \sum_{i}q^3_{t_{i}} + \sum_{i}q^3_{n_{i}}=0\label{ano3}
\; .
\end{eqnarray}
Following \cite{Chen:2006hn}, we parametrize the charges as
\begin{eqnarray}
q_{t_{1}}&=&-\frac{1}{3}q_{f_{1}}-2a \; ,\\
q_{t_{2}}&=&-\frac{1}{3}q_{f_{2}}+a+a' \; ,\\
q_{t_{3}}&=&-\frac{1}{3}q_{f_{3}}+a-a' \; ,
\end{eqnarray}
and
\begin{eqnarray}
q_{n_{1}}&=&-\frac{5}{3}q_{f_{1}}-2b \; ,\\
q_{n_{2}}&=&-\frac{5}{3}q_{f_{2}}+b+b' \; ,\\
q_{n_{3}}&=&-\frac{5}{3}q_{f_{3}}+b-b' \; .
\end{eqnarray}
With this parametrisation, the 
conditions (\ref{ano1}) and (\ref{ano2}) 
are satisfied automatically. The values of $q_{f_{i}}$, $a$,
$a^\prime$, $b$ and $b^{\prime}$ are constrained by the cubic
equation (\ref{ano3}), as well as the observed fermion masses and mixing
patterns.

\section{Fermion Masses and Mixings}

The up-type quark mass matrix is given by the Yukawa coupling,
\begin{equation}
\lambda^{|q_{t_{i}} + q_{t_{j}} + q_{H_{1}}|} {\bf
10}_{i}{\bf10}_{j} {\bf5}_{H_{1}} \; , \label{eq:up}
\end{equation}
where $\lambda =  \langle\phi\rangle / \Lambda$. We will take the
expansion parameter to be the Cabibbo angle, $\lambda\sim0.22$.
Note that if the sum of the charges
$( q_{t_{i}} + q_{t_{j}} + q_{H_{1}} )$ is non-integer for some $i$ and $j$,
that particular Yukawa coupling is forbidden.

In general, there are similar
operators involving ${\bf{5}}_{H_{2}}$ which contribute to
the up-type quark masses
and thus must be included.
As we will show later, due to the $U(1)_{F}$ charge of the ${\bf
5}_{H_{2}}$, these operators are suppressed because the sum of charges
$( q_{t_{i}} + q_{t_{j}} + q_{H_{2}} )$
is non-integer for all $i$ and $j$.
It is thus sufficient to consider only the  operators given in
Eq.~(\ref{eq:up}). 
In this paper, we restrict ourselves to 
the case with $( q_{t_{i}} + q_{t_{j}} + q_{H_{1}} )>0$. The
exponents that determine the quark mass matrix elements $U_{ij}$
are then
\begin{eqnarray}
\begin{pmatrix}
~|2q_{t_{1}}+q_{H_{1}}|~ & ~|q_{t_{1}}+q_{t_{2}}+q_{H_{1}}|~
& ~|q_{t_{1}}+q_{t_{3}}+q_{H_{1}}|~ \\
~ & |2q_{t_{2}}+q_{H_{1}}|
& |q_{t_{2}}+q_{t_{3}}+q_{H_{1}}| \\
&&|2q_{t_{3}}+q_{H_{1}}| \end{pmatrix}~
\end{eqnarray}
with $U_{ij} = U_{ji}$.

As the top quark mass is large, it is natural to assume that it is
un-suppressed by the expansion parameter. We therefore demand 
$2q_{t_{3}}+q_{H_{1}}=0$. We also assume $q_{f_{2}}=q_{f_{3}}$,
which is motivated by the large atmospheric neutrino mixing.
With these assumptions, the parameters  
\begin{equation}
a'=1,~~-\frac{1}{3}(q_{f_1}-q_{f_2})-3a=2 \; ,  \label{eq:up1}
\end{equation}
gives the up-type quark Yukawa couplings
\begin{eqnarray}\label{fu}
Y_u\sim\begin{pmatrix}\lambda^6&\lambda^5&\lambda^3\\\lambda^5&\lambda^4&
\lambda^2\\\lambda^3&\lambda^2&1\end{pmatrix} \; ,
\end{eqnarray}
yielding a realistic up-type quark mass
hierarchy~\cite{Sato:1997hv}.

Down-type quark masses are generated by the Yukawa couplings,
\begin{equation}
\lambda^{|q_{t_{i}} + q_{f_{j}} - q_{H_{2}}|}
{\bf10}_i{\bf\overline5}_j{\bf\overline5}_{H_{2}} \; .
\end{equation}
All couplings to the $\overline{5}_{H_{1}}$ are highly suppressed,
because the corresponding sums of the $U(1)_F$ charges 
are non-integer, as we shall see when we present the solutions for the
charges.

Let us assume that the b-quark mass is generated at the renormalisable
level. (In fact we have also examined the cases when the 
$b$-mass Yukawa is suppressed by a factor of $\lambda^{\alpha_b}$
with $\alpha_b  = 1,2,3$, 
but without finding a solution more elegant than the one we present here). 
The exponents of the elements in the down-type quark mass matrix
are then \begin{eqnarray} \begin{pmatrix}~|-9a-3| ~ &~ 3~ &~ 3~\\ ~
|-9a-4| ~&~2~&~2~\\~ |-9a-6| ~ &~0~&~0~\end{pmatrix} \; . \end{eqnarray}
The observed mass hierarchy among the down-type quarks can be obtained
with $a=-\frac{7}{9}$ and $q_{f_{1}}-q_{f_{2}}=1$ when we take
$(q_{t_{i}} + q_{f_{j}} - q_{H_{2}})>0$. The Yukawa couplings for the
down type quarks and the charged leptons are  \begin{eqnarray}\label{fd}
Y_d\sim
Y^T_e\sim\begin{pmatrix}\lambda^4&\lambda^3&\lambda^3\\\lambda^3&
\lambda^2&\lambda^2\\\lambda&1&1 \end{pmatrix} \; . \end{eqnarray} The
Yukawa matrices $Y_{u}$ and $Y_{d}$ also give rise to realistic CKM
matrix elements~\cite{Sato:1997hv}. In addition, the Georgi-Jarlskog
relations~\cite{Georgi:1979df} for the first and second generations of
down type quarks and charged leptons can be obtained by introducing a
$\bf 45$-dim Higgs (accompanied by a $\overline{{\bf 45}}$ to maintain 
anomaly cancellation).  The following discussion does not depend on this.

Let us return to the anomaly cancellation conditions. The cubic equation
in terms of the three free parameters $b$, $b^{\prime}$ and $q_{f_{2}}$
reduces to a linear one and it is given by \begin{equation} q_{f_2} = -
\frac{4550+2430b^{2} + 729b^{3} - 81 b (-25+9b^{\prime \, 2})}
{45(124+90b+81b^{2} + 27b^{\prime \, 2})} \; . \end{equation} Thus for
any rational values of $b$ and $b^{\prime}$, there always exists a
solution for  $q_{f_2}$. The simplest set of solutions we found
correspond to $b = -37/18$ and $b^{\prime} = 3/2$.  The corresponding
$U(1)_F$ charges for all the fields in the model are shown in Table I.
It is remarkable that such a simple solution to all the anomaly
cancellation  constraints exists. We note that there are degenerate
solutions. With $b=-5/9$ and $b^{\prime} = 3$, we again get $q_{f_2}= -1/2$, 
but now $q_{n_{1}} = 5/18$
and $q_{n_{2}}=59/18$. These charges lead to the same effective neutrino
mass matrix, $m_{\nu}$, as given in Eq.~(\ref{lightneutrinomass}) below.

We now consider the neutrino sector. 
The neutrino Dirac mass matrix is generated by the Yukawa
couplings 
\begin{equation}
\lambda^{|q_{f_{i}} + q_{n_{j}} + q_{H_{1}}|}
{{\bf\overline5}_i}N_j{{\bf5}_{H_{1}}} \; .
\end{equation}
As in the charged fermion sector, we will find that all couplings to 
the other Higgs (in this case ${\bf5}_{H_{2}}$)
are highly suppressed. 
With the charge assignment given in Table I, the neutrino Dirac Yukawa
matrix is given by
\begin{eqnarray}
Y_{\nu}\sim
\begin{pmatrix}
\lambda^{7} & \lambda^4 & \lambda\\ 
\lambda^{6} & \lambda^3 & 1\\
\lambda^{6} & \lambda^3 &1\end{pmatrix}
\; .
\end{eqnarray}

In order to use the type-I seesaw mechanism 
to generate the effective neutrino masses, we must introduce a second pair 
of SM-singlets $\chi, \chibar$ with charges $\mp 5/9$. Then the 
Majorana mass matrix for the right-handed neutrinos is
\begin{eqnarray}
M_{RR}
\sim\begin{pmatrix}
\lambda^{6} & \lambda^3 & 1\\
\lambda^{3} & 1 & \lambda^3 \\
1 & \lambda^{3} & \lambda^{6}
\end{pmatrix}
\left< \chi \right> \; .
\end{eqnarray}
The effective light neutrino mass matrix, after implementing the
seesaw mechanism, is
\begin{eqnarray}\label{lightneutrinomass}
m_{\nu}\sim
Y_{\nu}M_{RR}^{-1}Y^T_{\nu}v^2\sim
\begin{pmatrix}
\lambda^{8} & \lambda^{7} & \lambda^{7} \\
\lambda^{7} & \lambda^{6} & \lambda^{6} \\
\lambda^{7} & \lambda^{6} & \lambda^{6} \end{pmatrix}
\frac{v^2}{\left<\chi \right>} \; ,
\end{eqnarray}
where $v = \vev {H_1}$. If we assume $v\sim 240 \GeV$ and that the largest 
neutrino $(\hbox{mass})^2$ is around $2\times 10^{-3}\eV^2$, we find
$\left< \chi \right> \sim 10^{11} \;$ GeV. 
The textures given in Eqs.~(\ref{fu}), 
(\ref{fd}) and (\ref{lightneutrinomass}) have been shown to give successful
fermion masses and mixings~\cite{Sato:1997hv}, including those in
the neutrino sector~\cite{Chen:2007afa}.

%
%


\begin{table}
\begin{tabular}{|c|c| c| c| c| c| c| c|c|c|c|c|}
\hline Field&${\bf \overline{5}}_1$&${\bf \overline{5}}_2$&${\bf
\overline{5}}_3$&${\bf 10}_1$&${\bf 10}_2$&${\bf
10}_3$&$N_1$&$N_2$&$N_3$&${\bf 5}_{H1}$&${\bf 5}_{H2}$\\\hline
$U(1)_F$
charge&$\frac{1}{2}$&$-\frac{1}{2}$&$-\frac{1}{2}$&$\frac{25}{18}$&
$\frac{7}{18}$&$-\frac{29}{18}$&$\frac{59}{18}$&$\frac{5}{18}$
&$-\frac{49}{18}$&$\frac{29}{9}$&$-\frac{19}{9}$\\
\hline
\end{tabular}
\caption{$U(1)_F$ charges of different fields.}
\end{table}


\section{Proton decay}
The usual minimal SUSY $SU(5)$ GUT model suffers from the problem
of having rapid proton decay due to the dimension 5 operators
mediated by colored triplet Higgsinos, if the masses of the
SUSY particles are $\sim 1~{\rm TeV}$ \cite{protondecay}. 

In our model the R-parity violating operators 
$\lambda_{ijk}{\bf10}_i{\bf\overline5}_j{\bf\overline5}_k$
are forbidden by our $U(1)_F$ charge assignments, and moreover 
cannot be generated via higher dimensional terms involving 
powers of $\phi,\phibar$ and/or $\chi,\chibar$. 
(In fact with right handed neutrinos 
we also have the possibility of 
dimension-4 R-parity violating operators of the form 
$N_i N_j N_k$ and $N_i {\bf 5}_{H_{1,2}}{\bf\overline 5}_{H_{1,2}}$; these 
are similarly forbidden).

Let us now consider the dimension 5 operators $\kappa_{ijkl}{\bf10}_i{\bf10}_j
{\bf10}_k{\overline5}_l$. In the usual SUSY GUT theory, 
one needs to tune the
parameters $\kappa_{1121},\kappa_{1122}$ to be smaller than
$10^{-8}/M_{Pl}$ \cite{Hinchliffe:1992ad}.
These operators are also forbidden here by $U(1)_F$ conservation, but can 
be generated from higher-dimension operators involving $\phi,\phibar$ and 
$\chi,\chibar$.  for example the operator 
${\bf10}_1{\bf10}_1{\bf10}_2{\overline5}_1$ has $U(1)_F$ 
charge $11/3$, which can be generated from the operator
\begin{equation}
{\bf10}_1{\bf10}_1{\bf10}_2{\overline5}_1 
\left( \frac{\phi^2\chi^3}{\Lambda^6}\right)
\end{equation}
which for $\Lambda \sim M_{\hbox{\small GUT}}$ is very suppressed. All the 
$\kappa_{ijkl}$ operators are suppressed by $({\vev \chi}/\Lambda)^3$. 
\begin{figure}[t]
\includegraphics[scale=0.59]{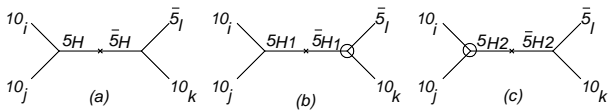}
\caption{Feynman diagrams of dimension 5 operators that lead to proton
decay mediated by color triplet Higgsinos. (a): Dimension 5 proton
decay operators in the usual minimal SUSY $SU(5)$
model~\cite{Sakai:1981pk}. (b) and (c): in the presence of the
$U(1)_{F}$, these operators are absent because the Yukawa
couplings that are circled are highly suppressed.}
\end{figure}

In the standard $SU(5)$ treatment these operators are generated by 
color triplet Higgs exchange. 
In our model, however, we have two conjugate pairs of Higgses, 
$({\bf 5}_{H_{1}} \oplus {\bf\overline 5}_{H_{1}}$) 
and $({\bf 5}_{H_{2}} \oplus {\bf\overline
5}_{H_{2}}$). With the $U(1)_F$ charge assignment given in Table I,
the couplings ${\bf10}_i{\bf10}_j{\bf5}_{H_{2}}$ and
${\bf10}_i{\bf\overline5}_j{\bf\overline5}_{H_{1}}$ are also suppressed 
by $({\vev \chi}/\Lambda)^3$ for any $i$ and $j$, 
because the sums of the $U(1)_F$ charges of the
fields involved in each of these operators are fractions of the form 
$2/3, -1/3, 7/3$ etc. 
The mass terms that mixes  ${\bf5}_{H_{1}}-{\bf\overline5}_{H_{2}}$ and  
${\bf5}_{H_{2}}-{\bf\overline5}_{H_{1}}$ are similarly suppressed by 
a factor of  $({\vev \chi}/\Lambda)^3\lambda^7$. 
In Fig.~(1) we contrast the situations in the standard $SU(5)$ treatment  
and in our model.

Therefore, all dangerous dimension 5 operators that could lead to fast proton decay
are absent in this model~\footnote{Of course it is possible to envisage models where dimension 5 proton
decay  contributions are permitted which are  nevertheless consistent
with observations~\cite{Berezhiani:2006mt};  however it seems to us more attractive if
they are  forbidden in a natural way.}.

We note that a similar mechanism to suppress dimension 5 proton decay
operators has been discussed in, for example~\cite{Babu:1993we,Dreiner:2007vp}.
For more recent work utilising a discrete symmetry, see 
\cite{Mohapatra:2007vd}. In these models, unlike in our
case, the cubic anomaly cancellation condition was not imposed
to constrain the charges.

\section{Conclusion} We have constructed a realistic model based on SUSY
$SU(5)\times U(1)_F$, which is free {\it all gauge anomalies}.  It is
quite remarkable that  we are able to find such a simple solution for
the charges that achieves this. Realistic
fermion masses and mixing angles are generated upon breaking of the 
$U(1)_{F}$ symmetry. We find that three right-handed neutrinos are
required in this model in order to cancel the gauge anomalies, in
addition to generating neutrino masses. Most interestingly,  all dimension 5
operators that could lead to proton decay are automatically suppressed.
The model therefore possesses  all the successes of grand unification, 
while still being consistent with the limits from non-observation of
proton decay.

We have not discussed the supersymmetry-breaking sector of the theory, 
nor the origin of the low energy Higgs potential. One might, for
example, consider  anomaly mediated supersymmetry-breaking; particularly
since  then introduction of an anomaly-free $U(1)$ has been advocated 
as leading to a solution of the  tachyonic slepton problem. However for
this to work all  the lepton doublets and the charged lepton singlets 
must have the same sign  of the $U(1)$ charge, so it is not compatible
with  the structure of our model.  In a non-GUT context a viable 
marriage of FN textures with anomaly mediation and a $U(1)_F$ was
described in  Ref.~\cite{Jack:2003qg} (although in that analysis, unlike
here, exotic  SM singlets are again required to cancel the $U(1)_F$
cubic and  gravitational anomalies). 

Since our quark and lepton mass matrices arise from Yukawa couplings  to
${\bf 5}_{H_1}$ and ${\bf\overline{5}}_{H_2}$, we need  the light Higgs
doublets $H_u$  and $H_d$ to come primarily from these representations.
Note that as indicated  above the
${\bf5}_{H_{1}}-{\bf\overline5}_{H_{2}}$ mass term  is suppressed;
however the ${\bf5}_{H_{1}}-{\bf\overline5}_{H_{1}}$ and
${\bf5}_{H_{2}}-{\bf\overline5}_{H_{2}}$ mass terms are allowed.  One
way to obtain light Higgs doublets would be to introduce  couplings
${\bf5}_{H_{1}}{\bf 24}\,{\bf\overline5}_{H_{1}}$ and
${\bf5}_{H_{2}}{\bf 24}\,{\bf\overline5}_{H_{2}}$ tuned as in the 
original supersymmetric $SU(5)$ model so as to leave  2 pairs of light
Higgs doublets and heavy Higgs triplets. Having two pairs of  Higgs
doublets would mean that there would have to be quite large  threshold
corrections in order to maintain gauge unification, unless  one arranged
to have a one light pair of color triplets. (This possibility  has been
considered recently in $E_6$-based models~\cite{Athron:2007en}.) We hope
to return elsewhere to a full construction of the  Higgs sector of the
theory.

It would be interesting to see if this model can be realized in a
more direct way, for example in a string theory model of
intersecting branes. 
It may be possible to 
implement the mechanism of \cite{Frampton:2002qc}, 
in which the connection between leptogenesis and low energy 
leptonic CP violation can be established~\cite{Chen:2004ww}. 
It would be interesting to investigate this further.


\section{Acknowledgements}
The authors thank K.T. Mahanthappa and R. N. Mohapatra for useful
discussions. The work of M.-C.C. and H.B.Y. is supported in part by the
National Science Foundation under Grant No. PHY-0709742. The work of
A.R. is supported in part by the National Science Foundation under Grant
Nos. PHY-0354993 and PHY-0653656.




\end{document}